# NIR and visible structured beam generation with a dual-force mechanical long-period fiber grating

**Wen-Hsuan Kuan[1,2,*], Xin-Yu Hou[1], Kuei-Huei Lin[1,*], and Vidar Gudmundsson[2]**

[1] Department of Applied Physics and Chemistry, University of Taipei, Taipei, Taiwan
[2] Science Institute, University of Iceland, Reykjavik, Iceland

*Authors to whom any correspondence should be addressed.

**E-mails:** wenhsuan.kuan@gmail.com, khlin@utaipei.edu.tw, isa.xyhou@gmail.com, vidar@hi.is

## Abstract

This work presents tunable generation of vortex, vector, and flat-top 1060-nm NIR beams in a few-mode fiber with a dual-force mechanical long-period fiber grating. By varying the applied forces on the fiber grating, the core mode to higher-order mode excitation can be adjusted. Manipulating the beam transformation is achieved by controlling the polarization of the fiber eigenmodes and the mode-coupling efficiency. In addition, our dual-force scheme enables greater flexibility in the mode selection by introducing the force gradient as an additional degree of freedom. By precisely tuning the intensity ratio between fundamental and doughnut modes, we arrive at the generation of propagation-invariant flat-top beams, for which the transverse intensity distribution exhibits a near-uniform profile over a propagation distance of 5 m in free space. The transverse optical field of 532-nm green light from a frequency-doubled Nd-doped yttrium vanadate laser can also be manipulated and coupled into various intensity distributions in a few-mode fiber by using a mechanically induced long-period fiber grating. We show that a doughnut beam, Mexican-hat beam, and crater-lake beam can be generated from the input Gaussian beam via the coupling of the fundamental core mode to a series of co-propagating higher-order modes with properly applied forces and polarizations. Our experimental results establish dual-force mechanically induced LPGs as a flexible and reconfigurable fiber-based tool for structured beam engineering, offering a unified platform for *in-situ* and robust control of phase, polarization, and amplitude of optical fields across different spectral bands.

## 1. Introduction

Unlike conventional Gaussian beams or plane waves, structured light features deliberately engineered transverse distributions of amplitude, phase, or polarization, enabling a wide range of applications. Among various forms of structured light, vector, vortex, and flat-top beams have attracted considerable interest due to their unique field structures and degrees of freedom [1-2].

From a modal perspective, these structured beams can be understood as arising from controlled superposition of fundamental and higher-order eigenmodes, either in free space or in guided-wave systems [3-4]. In particular, vector beams exhibit spatially varying polarization distributions, vortex beams carry orbital angular momentum associated with helical phase fronts, and optical fields combining polarization and phase inhomogeneities are commonly referred to as vector vortex beams [1]. For cylindrically symmetric media, both classes might display doughnut-shaped intensity profiles and singularities in their lowest-order forms.

Owing to these distinctive intensity, phase, and polarization properties, doughnut-shaped structured beams have found widespread applications in laser material processing, optical trapping and manipulation, and super-resolution microscopy, motivating the development of stable and flexible generation schemes in both free-space and fiber-based systems [5-6].

Typical methods for generating doughnut-shaped structured beams rely on bulk optical components, enabling flexible control of phase and polarization distributions, but at the cost of increased system complexity, stringent alignment requirements, and limited long-term stability. These free-space architectures are often bulky and environmentally sensitive, making them less favorable for applications that require robustness, compactness, or integration.

To overcome these limitations, fiber-based approaches have been explored as an alternative solution for structured beam generation. For instance, offset launching between single-mode and multi-mode fibers has been demonstrated to produce radially, azimuthally, or hybridly polarized beams. However, such schemes typically suffer from significant coupling losses and limited controllability over mode conversion [7]. As a result, achieving stable and tunable generation of structured beams in an all-fiber configuration remains a nontrivial challenge.

Beyond doughnut-shaped intensity profiles, flat-top beams with nearly uniform transverse intensity distributions are also of considerable interest due to their applications in laser material processing, lithography, and optical manipulation [8]. Conventional flat-top beams generated through Gaussian beam shaping, however, are not eigenmodes of free space and therefore exhibit pronounced diffraction-induced distortion during propagation. This intrinsic limitation motivates the search for propagation-invariant flat-top beams supported by well-defined mode structures.

Formed through the coherent superposition of modes with tailored polarization states, vector flat-top beams can mitigate diffraction effects and maintain a quasi-uniform intensity profile over extended propagation distances. Nevertheless, most reported implementations rely on free-space optical setups, including planar source synthesis, conical refraction in biaxial crystals, or polarization-dependent spatial light modulation [8], where the achievable beam quality and uniformity are highly sensitive to alignment and mode matching. These constraints further highlight the need for an alternative approach that combines mode controllability, propagation stability, and experimental simplicity within a guided-wave architecture.

Based on forward coupling between the fundamental core mode and co-propagating higher-order modes, long-period fiber gratings (LPGs) provide an effective all-fiber tool for mode engineering, enabling controlled manipulation of transverse intensity, phase, and polarization distributions in guided-wave systems [3-4]. In contrast to fiber Bragg gratings, which couple counter-propagating modes, LPGs operate through phase-matched forward coupling and are therefore well suited for mode conversion and structured beam generation.

Among various methods, mechanically induced LPGs are particularly attractive due to their tunability and reconfigurability, allowing the coupling characteristics to be continuously adjusted through external stress without permanent modification of the fiber structure [9]. By improving this approach, we previously demonstrated ultrabroadband mode conversion in few-mode fibers (FMFs) using dual-force mechanically induced LPGs, establishing a versatile and tunable scheme for guided-mode engineering rather than a fixed narrowband component [10]. An ultrabroadband long-period fiber grating with 10-dB bandwidth of 155 nm and 3-dB bandwidth of 415 nm is demonstrated around an optical wavelength of 1060 nm in a few-mode fiber. Combined with a wavelength-tunable ytterbium-doped fiber laser, the application of the dual-force method achieves the generation of doughnut beams in the spectral range of 1030 nm $\sim$ 1099 nm.

Building upon this capability, the present work explores the use of mechanically induced LPGs for controlled superposition of multiple guided modes and polarization states, enabling the tunable generation of a variety of structured light fields in optical fibers. Specifically, we demonstrate the generation of vortex beams, cylindrical vector beams (CVB), and propagation-invariant flat-top beams at a wavelength of 1060 nm in a few-mode fiber by jointly tuning the applied mechanical forces and the polarization states of the fiber eigenmodes. By precisely adjusting the coupling ratio between the fundamental and higher-order modes, a flat-top beam with a nearly uniform intensity distribution is maintained over a propagation distance exceeding 5 m.

Furthermore, the versatility of this approach is extended to the visible spectral regime. Using the same mechanically induced LPG technology, we manipulate the transverse optical field of 532-nm green light from a frequency-doubled Nd:$YVO_4$ laser and generate a range of structured intensity profiles, including doughnut, Mexican-hat, and crater-lake beams, through selective excitation and superposition of multiple higher-order modes in a few-mode fiber. These results demonstrate that mechanically induced LPGs provide a unified and flexible approach for structured light generation across different spectral bands, combining mode controllability, polarization tuning, and propagation stability within an all-fiber architecture.

## 2. Experimental setup

Figure 1 shows the experimental setup for structured beam generation in the near-infrared (NIR) region based on a dual-force mechanically induced LPG. A wavelength-tunable ytterbium-doped fiber laser (YDFL), using an ytterbium-doped fiber amplifier (YDFA) as the gain medium, provides the optical source in the 1-μm band. The laser output is launched into a single-mode fiber and subsequently coupled into a few-mode fiber through fusion splicing. The FMF used in this setup is Corning SMF-28.

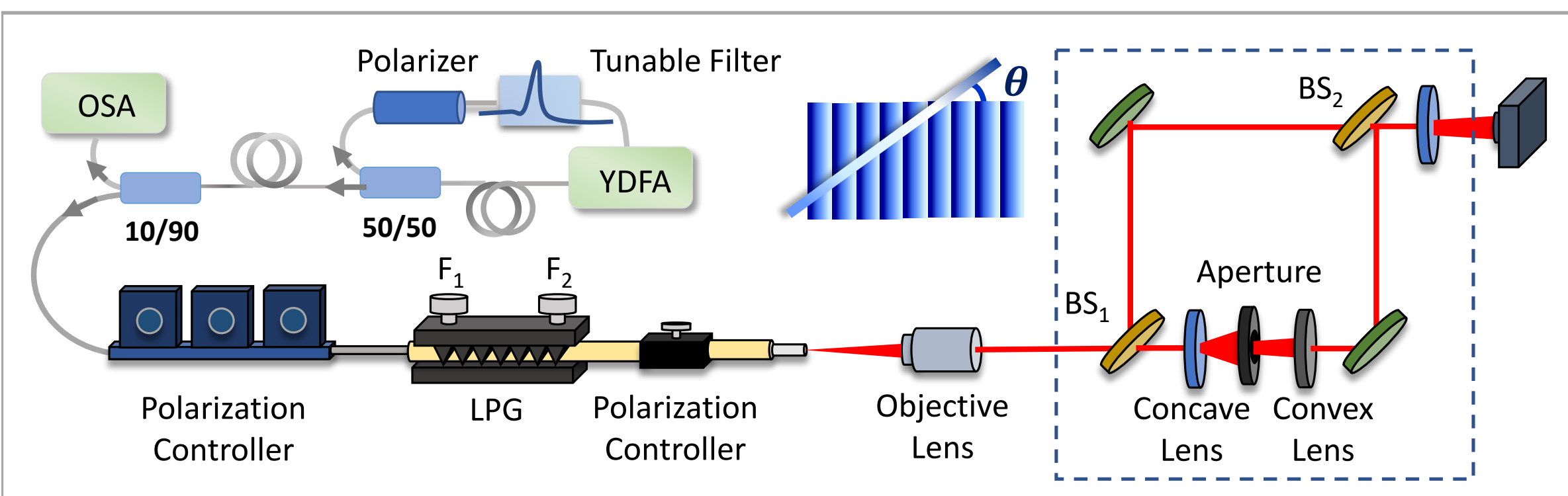

**Figure 1.** Experimental setup of tunable NIR beam generation and the Mach-Zehnder interferometer for the measurement of optical vortex beam.

A mechanically induced LPG is formed by sandwiching a section of the FMF between a periodic V-grooved plate and a flat plate. The period of grooves is 400 μm. By turning the two knobs, the forces $F_1$ and $F_2$ on both ends of the V-grooved plate can be different to introduce a force gradient along the fiber, while the angle between FMF and grooves can be adjusted [10]. The inset of Fig. 1 shows the optical fiber and the V-grooved plate at an angle of $\theta$. Quasi-periodic stress is applied along the fiber, and a proper force gradient and FMF angle is introduced to enable coupling between the fundamental and higher-order guided modes. Polarization controllers are placed before and after the LPG section to control the polarization states and relative phases of the coupled modes. The transverse intensity distribution at the FMF output is recorded using an optical beam profiler. When phase characterization is required, a Mach–Zehnder interferometer is employed, where a small portion of the wavefront taken from an aperture is used as the reference.

The experimental setup for visible-wavelength mode manipulation is schematically shown in Fig. 2. The mechanically induced LPG is implemented in a Corning HI 1060 fiber, which is an FMF in the visible band and is sandwiched between the identical homemade V-grooved and flat plates. A frequency-doubled Nd:$YVO_4$ laser operating at 532 nm is used as the light source. The green laser beam is coupled into the FMF through an objective lens. Polarization controllers are used to adjust the polarization states prior and subsequent to the LPG. The output beam from the FMF is characterized using an optical beam profiler to measure the intensity distribution for investigating the mode coupling behavior under different applied forces and polarization conditions.

To relate the mechanical tuning of the LPG to the observed mode coupling behavior in FMF, the effective grating period $\Lambda_{eff}$ can be expressed as

$$\Lambda_{eff} = \Lambda_0 / \cos\theta,$$

where $\Lambda_0$ is the period of V-grooves and $\theta$ is the angle between the fiber and the normal of grooves. The resonance wavelength $\lambda_{res}$ and $\Lambda_{eff}$ are related by the phase-matching condition:

$$\lambda_{res} = \Lambda_{eff}(n_{01} - n_{lm}),$$

where $n_{01}$ and $n_{lm}$ are the effective indices of the $LP_{01}$ and $LP_{lm}$ modes, respectively. Therefore, the resonance wavelength can be adjusted by tuning $\Lambda_{eff}$ or the effective index difference. By adjusting $\theta$, we vary $\Lambda_{eff}$ accordingly, and the resonance wavelength of LPG is tuned [11]. On the other hand, the effective index difference can be changed by photo-elastic or thermo-optic effects. In addition, the bandwidth and coupling efficiency of LPG can be tuned by the forces on the two ends of V-grooved plate. Conventional mechanical LPGs using a single constant force offer only one degree of freedom, thus the tuning for LPG bandwidth and coupling strength is inherently dependent. In contrast, our proposed dual-force configuration provides two independent degrees of freedom, i.e., the average force and the force gradient, which enables us to separately control the LPG bandwidth and the coupling strength.

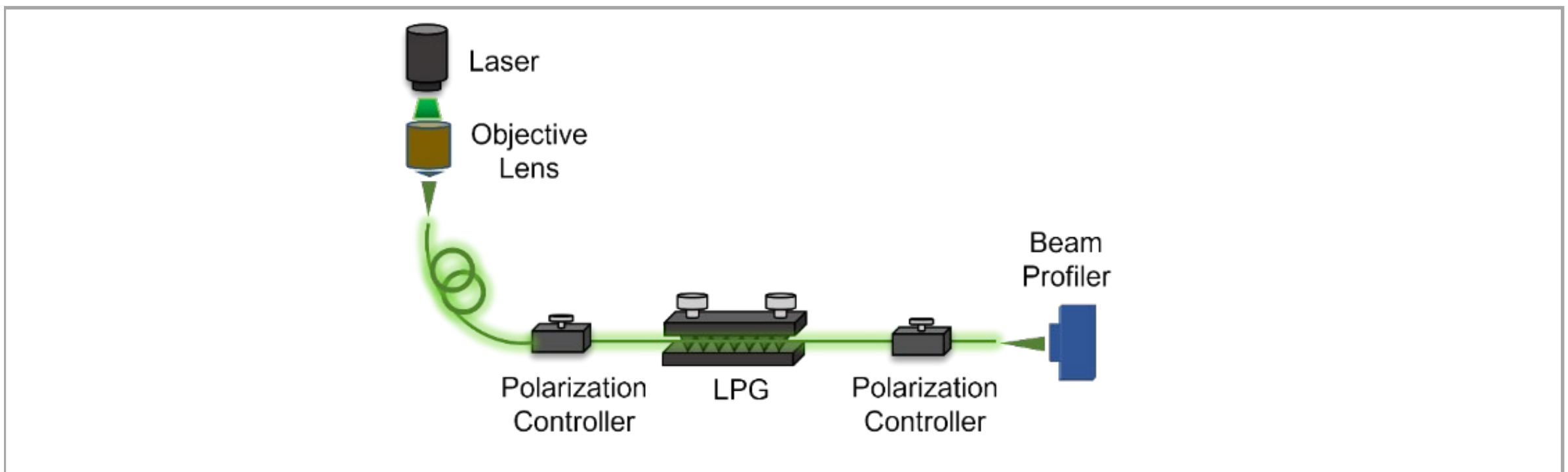

**Figure 2.** Schematic setup of few-mode LPG for the generation of various beams in the visible band.

## 3. Results and discussion

Figure 3 demonstrates the generation of optical vortex beams from the few-mode fiber using the dual-force mechanically induced LPG. A typical set of experimental parameters is $\theta$ = 27.8°, $\Lambda_{eff}$ = 452 μm, and the length of the LPG is 2.15 cm. The forces $F_1$ and $F_2$ are 13.72 N and 18.26 N, respectively. For Corning SMF-28 fiber, the nominal numerical aperture (NA) is 0.14 in broad wavelength range. At 1060 nm, the V parameter of Corning SMF-28 fiber is approximately 3.40, which is calculated using the nominal NA. Therefore, the fiber supports the propagation of two linearly polarized (LP) modes, i.e., $LP_{01}$ and $LP_{11}$. At 532 nm, the V parameter of Corning SMF-28 fiber is approximately 6.78, and the fiber supports seven LP modes ($LP_{01}$, $LP_{11}$, $LP_{21}$, $LP_{02}$, $LP_{31}$, $LP_{12}$, and $LP_{41}$).

The doughnut-shaped intensity profile in Fig. 3(a) shows the efficient excitation of $LP_{11}$ modes with LPG in the FMF. The observed interference characteristics in Fig. 3(b) and (c) confirm the formation of optical fields carrying well-defined orbital angular momentum (OAM) with opposite topological charges of $-\hbar$ and $+\hbar$, respectively. In this system, vortex beams are generated without the use of spatial light modulators or q-plates. Instead, the mechanically induced LPG enables efficient coupling from the fundamental $LP_{01}$ mode to the degenerate $LP_{11}$ mode group of the few-mode fiber. The mathematical expression for the normalized intensity profiles of the $LP_{01}$ and $LP_{11}$ modes in a step-index optical fiber is given by the square of their respective field solutions, utilizing Bessel functions $J_m$ within the core and modified Bessel functions $K_m$ in the cladding. A coherent superposition of the orthogonal spatial intensity modes $LP_{11a}$ and $LP_{11b}$ with equal amplitudes and a relative phase difference of $\pm\pi/2$ naturally gives rise to vortex states with $J_1$ core mode and opposite OAM, i.e., $OAM_{\pm} = LP_{11a} \pm i\, LP_{11b}$.

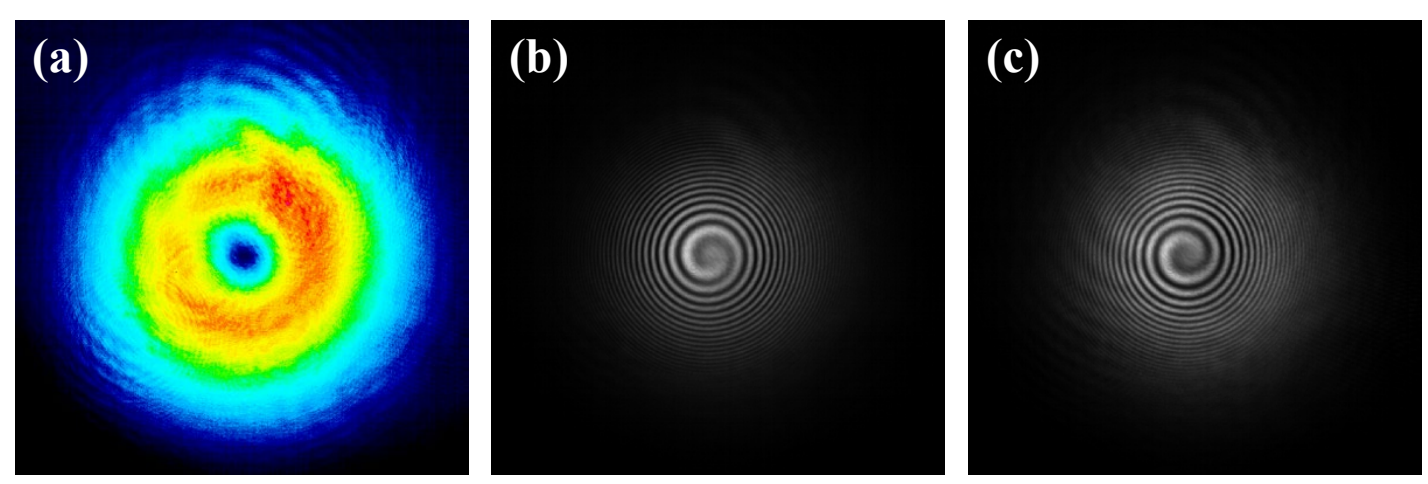

**Figure 3.** (a) The intensity profile of vortex beam generated from the few-mode fiber. In (b) and (c), the spiral interference patterns of the vortex beam with a reference beam show the OAM of $-\hbar$ and $+\hbar$, respectively.

The relative phase between the $LP_{11}$ mode components is controlled by polarization controllers in the fiber system, allowing continuous switching of the handedness of the helical phase front. By exploiting the intrinsic eigenmodes of the fiber and their mechanically tunable coupling, this all-fiber approach provides a compact and alignment-free method for vortex beam generation. The same mode-based phase control also forms the foundation for the generation of CVB and flat-top beams discussed in the following.

When the $LP_{11a}$ and $LP_{11b}$ modes are in phase or out of phase but with perpendicular linear polarizations, the coherent superposition will instead lead to the generation of CVB:

$$\text{Radially polarized CVB} = LP^{x}_{11a} + LP^{y}_{11b}$$

$$\text{Azimuthally polarized CVB} = LP^{y}_{11a} + LP^{x}_{11b}.$$

Here the superscripts $x$ and $y$ represent horizontal and vertical polarizations, respectively. Figure 4(a) shows the intensity profile of a CVB while (b)-(e) show the intensity profiles of an azimuthally polarized vector beam after passing through a linear polarizer with the transmission axis aligned clockwise at (b) 0°, (c) 45°, (d) 90°, and (e) 135° with the vertical direction, respectively. The result reveals the polarization-dependent response of an optical beam under linear polarization analysis, confirming its vector nature and the presence of spatially varying polarization. In our setup, radially polarized or hybridly polarized vector beams can also be achieved, depending on the adjustments of polarization controllers.

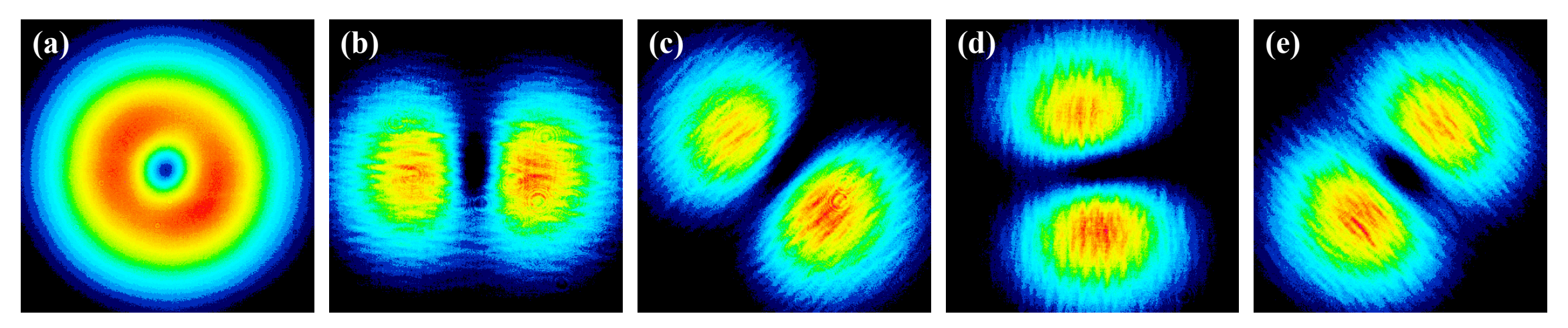

**Figure 4.** The intensity profiles of an azimuthally polarized vector beam (a) and after passing the vector beam through a linear polarizer with the transmission axis aligned clockwise at (b) 0°, (c) 45°, (d) 90°, and (e) 135° with the vertical direction.

By configurations of equal peak intensity weighting between the fundamental mode $LP_{01}$ and the higher-order mode $LP_{11}$, a flat-top beam profile is achieved via coherent mode superposition, as demonstrated in Fig. 5. The intensity profiles of fundamental mode, higher-order mode, and flat-top beam generated from the FMF are shown in Figs. 5(a), (b), and (c). Figures 5(d) and (e) depict the measured distributions and the curve fittings of optical intensity along the *x*-(horizontal) and *y*-axis (vertical) respectively with 4th-order super-Gaussian functions. The resulting transverse intensity distribution exhibits a near-uniform profile that remains stable over a propagation distance exceeding 5 m in free space, indicating propagation-invariant behavior. Unlike conventional free-space flat-top beams, which are not supported by propagation eigenmodes and therefore undergo significant diffraction-induced distortion, the flat-top beam observed here is sustained by fiber eigenmodes, leading to enhanced robustness during propagation [8]. The superposition of eigenmodes in our setup benefits from the coaxial propagation of the $LP_{01}$ and $LP_{11}$ modes in the FMF, as well as the simultaneous transformation of both modes by the objective lens or additional optics. The intensity profile of the superposition maintains its flat-top shape during propagation, albeit with a change in the beam size.

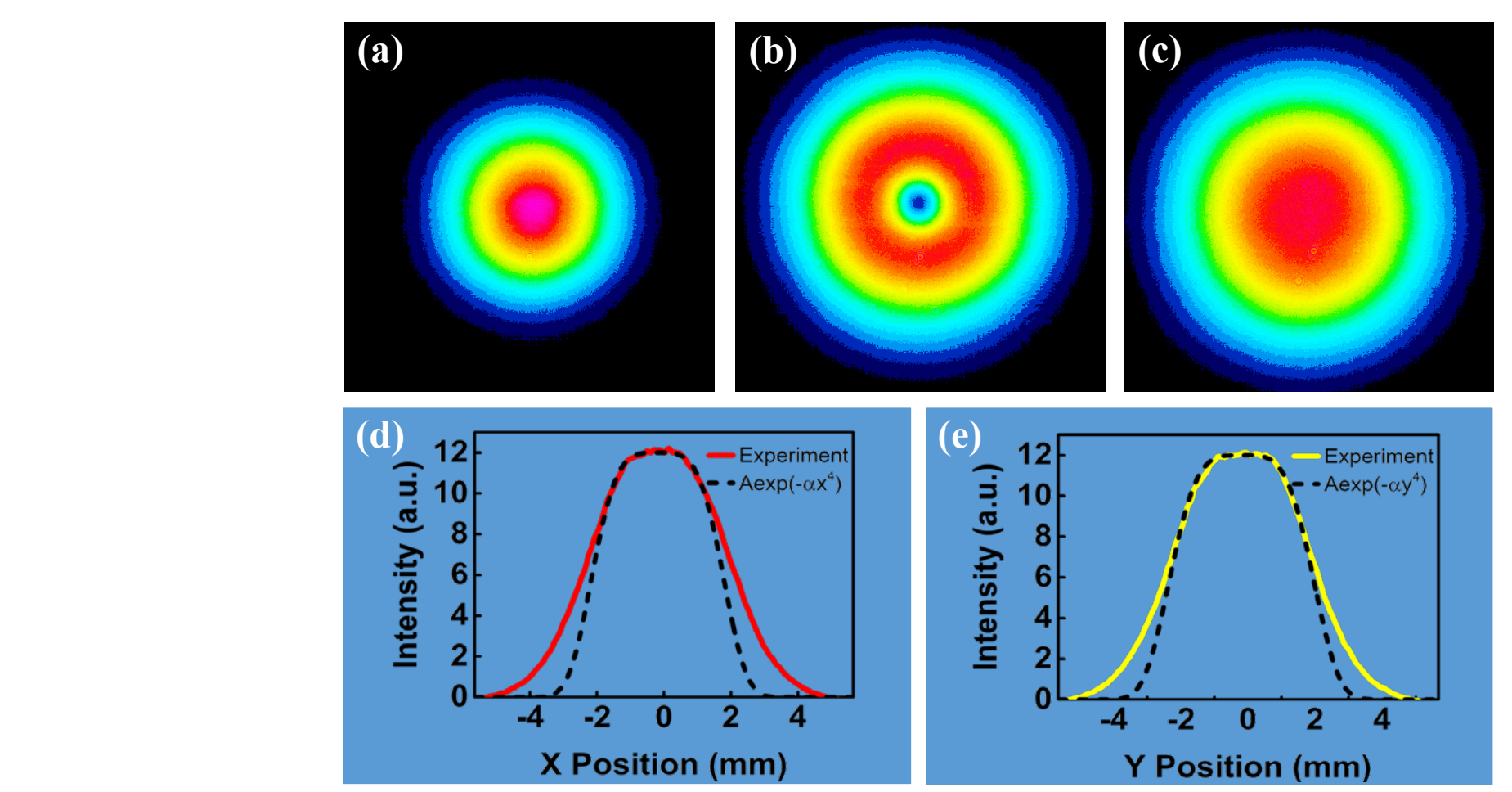

**Figure 5.** The intensity profiles of (a) fundamental mode, (b) higher-order mode, and (c) flat-top beam generated from the few-mode fiber. (d)-(e) The distributions of optical intensity along the *x*- and *y*-axis respectively.

Compared with free-space flat-top beam generation schemes reported in the literature [8], which typically rely on bulk beam-shaping elements, the implementation demonstrated here provides a compact and mechanically stable alternative by forming flat-top intensity profiles directly within a few-mode fiber through controlled mode superposition. Propagation-invariant flat-top beams are desirable for applications requiring uniform optical intensities, such as optical tweezers arrays and cold-atom platforms, where homogeneous illumination enables identical trapping conditions and reduced site-to-site inhomogeneity [12,13].

From a broader physical perspective, controlling guided modes via spatial or depth modulations in classical systems shares an underlying physical principle with state manipulation in quantum mechanics. This universal mechanism has been demonstrated in recent works on etched silicon splitters [14] and acoustic waveguides [15], as well as in laser-inscribed fiber gratings utilizing triple-superposition [16] or over-coupled states [17]. Furthermore, by introducing a linear period chirp, chirped long-period fiber gratings continuously vary the local phase-matching condition along the fiber axis [18]. This spatial variation provides a classical analogue of adiabatic rapid passage, in which the local phase mismatch acts as an effective detuning and drives smooth and nearly complete energy transfer between core modes through an avoided crossing of their effective propagation constants. Actually, by adjusting the force gradient, our system enables the variation in the force-induced effective index difference along the fiber. This allows us to successfully generate donut beams and achieve the adiabatic mode conversion. A related control principle appears in our previous quantum studies. In semiconductor quantum rings, chirped laser pulses sweep the optical detuning through resonance and induce robust cascaded population inversion [19]. In Bose–Einstein quantum droplets, chirping the relative frequency of the interfering laser fields produces an accelerating optical lattice, which sweeps the condensate quasi-momentum across an avoided band crossing and gives rise to nonlinear and nonreciprocal Landau–Zener transitions [20]. Together, these systems illustrate how spatially or temporally swept control parameters, including structural and frequency chirps, can govern classical mode conversion and quantum-state transfer through adiabatic and diabatic dynamics.

The generation of high-quality broadband structured beams in two-mode fibers requires precise control over mode coupling, yet conventional architectures like traditional LPGs remain permanently fixed once fabricated, which makes it extremely difficult to achieve an ideal coupling ratio for generating flat-top beam due to manufacturing tolerances and wavelength-dependent variations. Traditional laser-inscribed gratings rely on fixed, photo-induced periodic refractive index modulation, which cannot be altered post-fabrication. To overcome these limitations, we propose a flexible method utilizing a mechanical V-groove to induce periodic microbending and modulated refractive indices. Unlike fixed gratings, this mechanical approach enables real-time, *in-situ* adjustment of the coupling conditions, such as the force profiles, fiber tilt angles, and polarization states, during operation. By dynamically tuning the applied pressure, this technique provides reconfigurable control over individual mode weightings that effectively compensates for any non-uniformity in the coupling ratio, offering a robust and adaptable solution for correcting fabrication errors and ensuring precise structured beam generation with complete energy and state transfer across a broad spectrum.

The dual-force few-mode LPG can exhibit an ultrabroad coupling bandwidth in spite of the constant grating period, which is attributed to the variations of the propagation constant and of the effective refractive index difference of $LP_{01}$ and $LP_{11}$ modes along the FMF caused by the force gradient. Typical effective index difference between the $LP_{01}$ and $LP_{11}$ modes is estimated to be from $2.17 \times 10^{-3}$ to $2.51 \times 10^{-3}$ in the spectral range from 980 nm to 1135 nm [10]. The effective index difference obtained in our dual-force configuration is relative large as compared with the traditional LPG inscribed by $CO_2$ or UV lasers, which is beneficial for efficient mode coupling within a short section of LPG. It can also be utilized in the broadband mode converter based on over-coupled long-period fiber grating such as in Ref. 17. The dual-force few-mode LPG proposed in this work can thus provide effectively chirped and large effective refractive index difference for potential applications in broadband mode-division multiplexing, tunable optical filters, fiber lasers, structured beam generation, and other related fields. When used as an all-fiber optical filter, the center wavelength, the operation bandwidth, and the coupling strength of this LPG can be adjusted by the angle between the fiber and the grooved plate, the force gradient, and the average force, respectively. In addition to conventional fibers, the mechanical LPG technology can also be utilized for the photonic crystal fibers [11].

To examine the versality of this mechanical force-gradient technique, the idea is facilitated to the visible structured beam generation with 532-nm green light excitation. At the wavelength of 532 nm, the nominal NA of HI 1060 fiber is 0.14 and the V parameter is approximately 4.38. Therefore, the HI 1060 fiber have four LP modes ($LP_{01}$, $LP_{11}$, $LP_{21}$, and $LP_{02}$) at 532 nm. Measured intensity profiles of the fundamental beam, the doughnut beam, the Mexican-hat beam, and the crater-lake beam are shown in Fig. 6(a)-(d), while the simulation results are shown in (e)-(h). We compare experimentally observed annular beam profiles with representative numerical simulations, highlighting the role of higher-order mode superposition in shaping complex intensity distributions.

While EH and HE (demoted as $EH_{mn}$ and $HE_{mn}$) modes are the exact vector eigenmodes of a circular optical fiber, they reduce to $LP_{lm}$ modes under the weakly guiding approximation. The propagation constants of the vector eigenmodes are obtained by solving the dispersion relation based on Bessel functions and their first derivatives,

$$\left(\frac{J'_m(X)}{XJ_m(X)}+\frac{K'_m(Y)}{YK_m(Y)}\right)\left(\frac{n_1^2J'_m(X)}{XJ_m(X)}+\frac{n_2^2K'_m(Y)}{YK_m(Y)}\right)=\frac{m^2\beta^2}{k_0^2}\left(\frac{1}{X^2}+\frac{1}{Y^2}\right)^2,$$

where $X = ak_t = a\sqrt{k_0^2n_1^2-\beta^2}, Y=\sqrt{V^2-X^2}$, and $V = k_0a\sqrt{n_1^2-n_2^2}$, with the core radius of $a$, and the refractive indices of core and cladding being $n_1$ and $n_2$, respectively. For a given frequency $\omega = k_0c$ and fiber parameters $a, n_1$, and $n_2$, guided solutions exist only for $X < V$ and are determined by the intersections of the $J'_m(X)/[XJ_m(X)]$ curves corresponding to the TM and TE modes. All simulations are performed by the superposition of $LP_{01}$ and several higher-order $LP_{lm}$ modes, where the $HE_{11}$ cutoff is set by the first zero of $J_0(X)$. The fundamental mode $LP_{01}$ shown in Fig. 6(a) is constructed by synthesizing the $x/y$-polarized $HE_{11}$ field from the azimuthal component $E_\phi$ of the TM/TE solutions. In the absence of polarization-resolved measurements, the doughnut beam in Fig. 6(b) is simulated using the intensity distribution of the $TM_{01}$ mode. The increasingly complex radial structures observed in the Mexican-hat (Fig. 6(c)) and crater-lake beams [Fig. 6(d)] cannot be reproduced by a single guided mode, but instead arise from controlled superposition of low-order and higher-order guided modes. In the simulations, the Mexican-hat beam is obtained by combining the fundamental $HE_{11}$ mode with the $TM_{01}$ doughnut mode and additional higher-order modes from the $LP_{21}$ family, producing a central intensity dip surrounded by an annular ridge through appropriate modal weighting. The crater-lake beam is constructed from the same set of guided modes, but with a different balance among the modal weights, resulting in a distinct radial morphology with a central plateau and a surrounding ring.

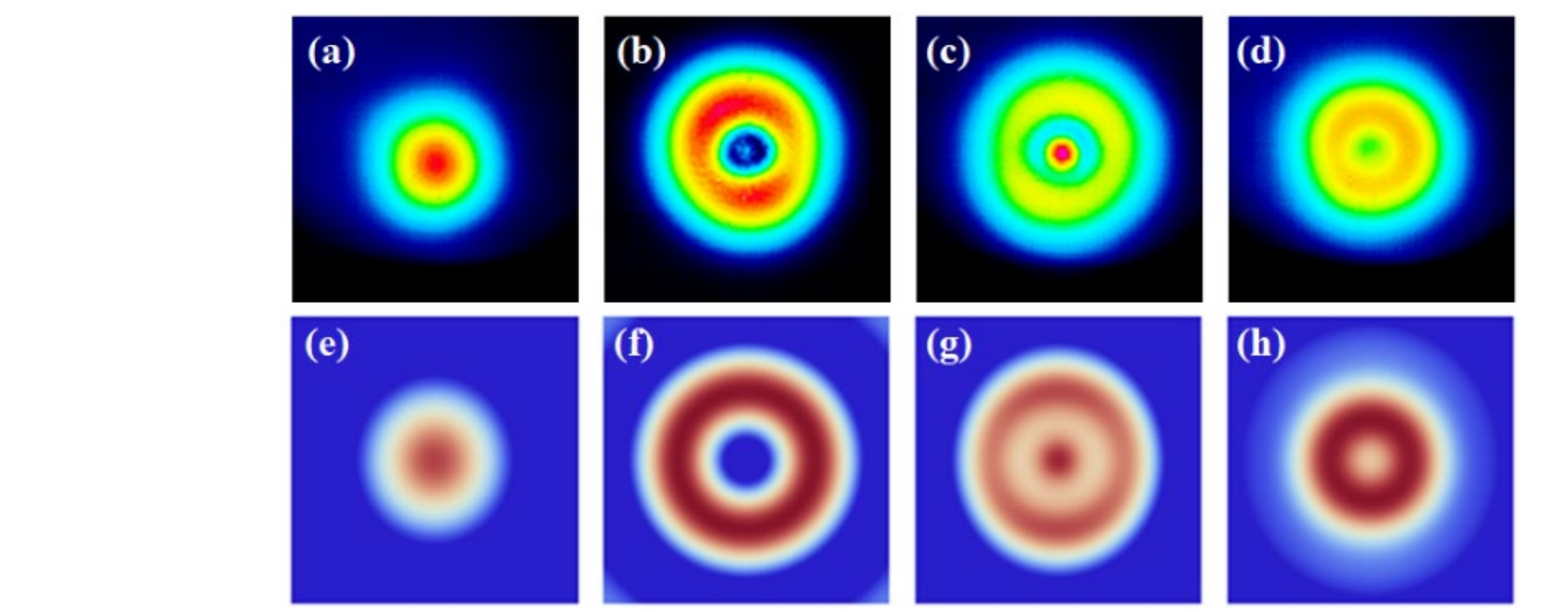

**Figure 6.** Intensity profiles of the fundamental beam, the doughnut beam, the Mexican-hat beam, and the crater-lake beam. (a)-(d): experimental measurements and (e)-(h): simulation results.

Compared with the near-infrared results, the visible-wavelength experiments further highlight the wavelength scalability and versatility of the mechanically induced LPG approach. The ability to access a richer variety of transverse intensity profiles at shorter wavelengths underscores the generality of guided-mode superposition as an effective strategy for structured light generation in fiber-based systems.

It's worth of emphasis that the complex transverse intensity profiles generated here, including Mexican-hat and crater-lake distributions, may provide reconfigurable optical potential landscapes for controlling atomic, molecular, and solid-state quantum systems. In cold-atom platforms, structured Gaussian and Laguerre–Gaussian fields have been employed to tailor light-induced fictitious magnetic trapping potentials [21]. Structured illumination has also been used to inscribe persistent lateral electrostatic superlattices in semiconductor heterostructures containing two-dimensional electron gases [22-23], while twisted light can transfer its linear and orbital angular momentum to the center-of-mass motion of optically generated excitons [24]. Superior to the traditional Gaussian beam, the mechanically reconfigurable LPG demonstrated here enables *in-situ* adjustment of the central depression, annular barrier, and outer radial slope of the optical intensity distribution. With appropriate light–matter couplings, this additional spatial degree of freedom could be explored for shaping bound-to-continuum wavepacket dynamics, modifying photoelectron angular distributions, or controlling molecular dissociation and semiconductor excitations.

## 4. Conclusions

In this paper, we have demonstrated an all-fiber system for generating a variety of structured light fields based on a dual-force mechanically induced long-period fiber grating. As compared with conventional mechanical LPGs using a single constant force, our proposed dual-force configuration provides two independent degrees of freedom (the average force and the force gradient), enabling us to separately control the LPG bandwidth and the coupling strength. By exploiting tunable mode coupling and polarization control in a two-mode fiber (SMF-28), vortex beams, CVB, and propagation-invariant flat-top beams are realized at a wavelength of 1060 nm without the use of free-space phase-modulating elements. The flat-top beam demonstrated here is supported by coherent superposition of guided fiber eigenmodes, resulting in a uniform transverse intensity distribution that remains stable over a propagation distance exceeding 5 m in free space. Compared with conventional free-space flat-top beam shaping techniques, this eigenmode-based approach provides enhanced robustness and stability within a compact guided-wave architecture.

Furthermore, the same mechanically induced LPG approach is shown to be effective in the visible spectral region with a four-mode fiber (HI 1060). The generation of Mexican-hat and crater-lake intensity profiles at 532 nm, together with supporting numerical simulations, confirms the versatility and wavelength scalability of guided-mode superposition for structured light generation. Even higher-order modes were successfully generated using the SMF-28 fiber (seven modes at 532 nm), though they are not shown here. These results verify that the mechanically adjustable LPG method can be effectively applied to few-mode fibers supporting a larger number of modes.

## Acknowledgements

We thank the National Science and Technology Council, Taiwan for partial financial support under grants NSTC 113-2221-E-845-003, NSTC 114-2221-E-845-002, and NSTC 115-2918-I-845-003. W.H.K thanks for the hospitality of the Science Institute of the University of Iceland.